\documentclass{aa}  
\usepackage{graphicx}
\usepackage{txfonts}

\newcommand\E[1]{\times10^{#1}}
\newcommand\avg[1]{\langle #1\rangle}
\newcommand\prob{{\cal P}}

\newcommand\U[1]{{\,\rm #1}}
\newcommand\Ub[1]{{\rm(#1)}}

\newcommand\surfbrig{W\,m^{-2}Hz^{-1}sr^{-1}}
\newcommand\flux{erg\,cm^{-2}s^{-1}}

\newcommand\al{\alpha}

\newcommand\eps{\epsilon}

\newcommand\lmb{\lambda}
\newcommand\sg{\sigma}

\newcommand\Sg{\Sigma}
\newcommand\Lmb{\Lambda}

\newcommand\SgD{$\Sigma$--$D$}

\newcommand\rs[1]{_\mathrm{#1}}

\newcommand\nz{n\rs{o}}
\newcommand\ESN{E\rs{SN}}
\newcommand\Vsh{V\rs{sh}}
\newcommand\epsB{\eps_B}
\newcommand\epsCR{\eps_{CR}}
\newcommand\Msun{M_\odot}
\newcommand\LX{L\rs{X}}

\newcommand\tP{\tilde\prob}

\begin{document}
   \title{A statistical approach to radio emission from shell-type SNRs}

   \subtitle{I. Basic ideas, techniques, and first results}

   \author{R. Bandiera\inst{1} \and O. Petruk\inst{2}}

   \offprints{R. Bandiera}

   \institute{INAF - Osservatorio Astrofisico di Arcetri
              Largo E. Fermi 5, I-50125 Firenze, Italy\\
              \email{bandiera@arcetri.astro.it}
        \and
              Institute for Applied Problems in Mechanics and Mathematics
              Naukova St. 3-b, Lviv 79060, Ukraine}

   \authorrunning{R. Bandiera \& O. Petruk}

   \titlerunning{A statistical approach to radio emission from shell-type SNRs}

   \date{Received 1 April 2009 / accepted 22 September 2009}

 
  \abstract
   {
Shell-type supernova remnants (SNRs) exhibit correlations between
radio surface brightness, SNR diameter, and ambient medium density,
that between the first two quantities being the well known $\Sigma$--$D$
relation.
   }
   {
We investigate these correlations, to extract useful
information about the typical evolutionary stage of radio SNRs, as well as to
obtain insight into the origin of the relativistic electrons and magnetic fields
responsible for the synchrotron emission observed in radio.
   }
   {
We propose a scenario, according to which the observed correlations are the
combined effect of SNRs evolving in a wide range of ambient conditions,
rather than the evolutionary track of a ``typical'' SNR.
We then develop a parametric approach to interpret the statistical data,
and apply it to the data sample previously published by Berkhuijsen, as
well as to a sample of SNRs in the galaxy M~33.
   }
   {
We find that SNRs cease to emit effectively in radio at a stage near the end of
their Sedov evolution, and that models of synchrotron emission with constant
efficiencies in particle acceleration and magnetic field amplification do
not provide a close match to the data.
We discuss the problem of the cumulative distribution in size, showing that
the slope of this distribution does not relate to the expansion law of SNRs,
as usually assumed, but only to the ambient density distribution.
This solves a long-standing paradox: the almost linear cumulative distribution
of SNRs led several authors to conclude that these SNRs are still in free
expansion, which also implies very low ambient densities.
Within this framework, we discuss the case of the starburst galaxy M82.
   }
   {
Statistical properties of SNR samples may be used to shed light on both the
physics of electron acceleration and the evolution of SNRs.
More precise results could be obtained by combining data of several surveys
of SNRs in nearby galaxies.
   }


   \maketitle
%

\section{Introduction}
\label{sect:Introd}

\indent
Radio emission is a quite common property of shell-type supernova remnants
(SNRs).
The intensity of the (synchrotron) radio emission is related to the magnetic
field strength and the amount of accelerated electrons.
However, the mechanisms leading to both the magnetic field amplification and
the electron injection at the SNR shock and their respective efficiencies
remain poorly constrained.
To investigate these processes observationally,
there have been detailed studies, mostly in X-rays, of some selected
SNRs (see e.g., Cassam-Chena\"\i\ et al.\ \cite{cea07}, Cassam-Chena\"\i\
et al.\ \cite{cea08}, and references therein).
Relevant, complementary information should also be extracted from a statistical
analysis of SNR data samples.

On the observational side, a well-known (even though not widely accepted)
statistical relation is the so-called ``\SgD\ relation'', namely
the empirical correlation discovered between the SNR size ($D$) and its
radio surface brightness ($\Sg$).
Various authors (e.g., Clark \& Caswell \cite{cc76}, Milne \cite{m79},
Caswell \& Lerche \cite{cl79}, Case \& Bhattacharya \cite{cb98}, Uro\v sevi\'c
et al.\ \cite{uea05}) have investigated this correlation.
Originally applied as a tool for estimating SNR distances, it has
been found to be unreliable for this purpose.
Some authors (e.g., Green \cite{g05}, Uro\v sevi\'c et al.\ \cite{uea05})
have also argued that this correlation may be affected by
selection effects.
For Galactic SNRs, Green (\cite{g05}) highlighted the selection effects:
(1) in the surface brightness, with a completeness limit of about
$10^{-20}\U{\surfbrig}$, while SNRs below this limit are predominantly in
regions where the Galactic background is low;
(2) in the angular size, so that young but distant SNRs may be missed.
For SNRs in other galaxies, Uro\v sevi\'c et al.\ (\cite{uea05}) also showed
that there may be
a selection effect in the integrated flux (valid for unresolved or mildly
resolved SNRs), with the effect of leading to an observed slope shallower
than the intrinsic one.

Even though selection effects may affect this correlation to some level,
we believe that the relation itself has a physical origin, and
that one can therefore extract from it information about the processes
involved in the
SNR radio emission: therefore, understanding its origin could eventually
contribute to constraining the efficiency of these processes.
There have been several attempts (e.g., Shklovsky \cite{s60}, Van der Laan
\cite{vdl62}, Poveda \& Woltjer \cite{pw68}, Kesteven \cite{k68}, or more
recently Duric \& Seaquist \cite{ds86}, Berezhko \& V\"olk \cite{bv04})
to explain the \SgD\ relation as the average evolutionary track of a
``typical'' SNR.
In all of these cases, the slope of the correlation is assumed to correspond
to that of the SNR evolutionary track on the \SgD\ parameter plane.

However, there is clear evidence that this assumption is incorrect.
Berkhuijsen (\cite{b86}) has found tight correlations of both
$\Sg$ and $D$ with the ambient density ($\nz$), in the sense that smaller
(and brighter) SNRs are typically located in a denser medium (indeed, the
correlation between $\nz$ and $D$ had already been known for several years;
see e.g., Fig.~4 in McKee \& Ostriker \cite{mo77}).
The best-fit results given by Berkhuijsen are
\begin{eqnarray}
\label{eq:DBerk}
D&\simeq&15\,\nz^{-0.39\pm0.04}\U{pc}					\\
\label{eq:SgBerk}
\Sg&\simeq&6\E{-20}\nz^{1.37\pm0.21}\U{\surfbrig}
\end{eqnarray}
Berkhuijsen concluded that the \SgD\ relation ($\Sg\propto D^\xi$) is
just a secondary effect, while the two primary relations are those of $D$
and $\Sigma$ with $\nz$.

\begin{figure}
\resizebox{\hsize}{!}{\includegraphics{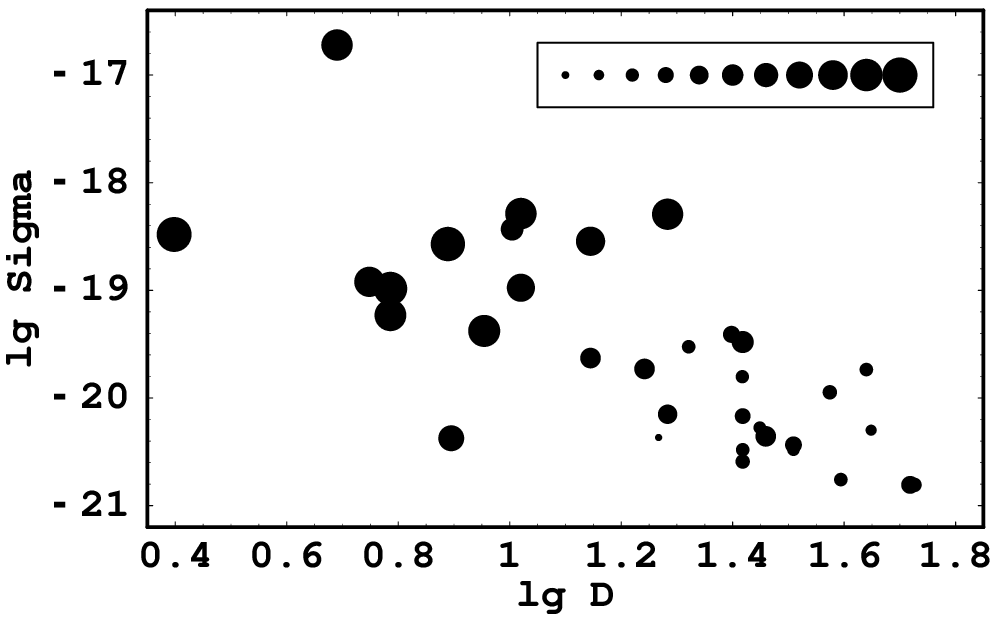}}
\caption{Distribution of SNRs from Berkhuijsen (\cite{b86}) 
sample in the $\lg
D$--$\lg\Sg$ parameter plane ($D$ is measured in pc, while $\Sg$ is in
$\U{\surfbrig}$).
The dot sizes are proportional to $\lg\nz$ values, the legend showing in
order sizes corresponding to $\lg\nz$ from --1.5 to 1.0, in steps of 0.25
($\nz$ being measured in $\U{cm^{-3}}$).
It is apparent from this figure that smaller (and brighter) SNRs are typically
located in a denser medium.}
\label{Fig1}
\end{figure}

Although this conclusion may appear rather extreme, it is quite obvious that
the \SgD\ relation contains SNRs that evolve in very different ambient
conditions.
We then assume that the correlations between $\Sg$, $D$, and $\nz$ do not
directly reflect the evolution of a ``typical'' object, but are rather the
combined effect of the evolution of different SNRs expanding in different
ambient conditions.
That the correlations with $\nz$ are statistically so tight also
suggests that $D$ and $\Sg$ are more sensitive to the ambient
conditions than to the SNR evolution.
The combined effect of these correlations is that a clear trend of $\nz$ across
the \SgD\ relation is observed (see Fig.~\ref{Fig1}), namely that smaller SNRs
are preferentially located in higher-density environments.

In principle, correlations of $\Sg$ and $D$ with $\nz$, accounting for at least
qualitatively the trend of $\nz$ across \SgD\ relation, could also appear
in the case of the evolution of an individual object, provided that it
expands in a medium with a highly structured, fractal, density distribution.
This would reflect that, during its life, a SNR always preferentially
expands towards the direction in which the ambient density is lower.
Therefore, at any time, the ``effective'' ambient density would be close to
the ``lowest'' ambient density in the volume occupied by the SNR.
In this way, one could explain why large SNRs typically seem to expand in a
low-density medium.

However, it is hard to justify, in this scenario, the absence of small SNRs
in low-density media (which would be the case, when the supernova itself is
located in a low density region), as well as that low ambient density values
are measured for all extended SNRs (which requires that such low density
regions are ubiquitous in the Galaxy, on scales of tens of parsecs or
even less).
In addition, if the effects of the fractal interstellar medium were dominant,
virtually all SNRs should have a much brighter limb on one side, which
is not observed.
Therefore, for all these reasons, we conclude that the ``fractal ambient
density'' hypothesis is implausible, and we do not consider it any longer
in this paper.

One of our goals is to show that the best-fit line usually referred to as the
``\SgD\ relation'' provides only a minor part of the information present
in the data, while additional information could be extracted by analyzing in
detail the distribution of points in the $\Sg$--$D$--$\nz$ parameter space.
For this reason we propose a rather general (parametric) scenario,
with the aim of constraining the physics of the electron
injection and magnetic field behavior in SNR shocks and/or the SNR
evolutionary phase in which they are most likely to be observed in radio.

The plan of the paper is the following.
In Sect.~2, we present the basic ideas, assumptions and formulae that will
be used in the rest of the paper.
Section 3 is devoted to a re-analysis of the data of Berkhuijsen (\cite{b86}),
using the criteria introduced in Sect.\ 2, and extracting constraints
on the dependence of the particle and magnetic field efficiencies on the
basic shock parameters.
For the sake of comparison, we also apply the same technique to a sample of
SNRs in galaxy M~33.
Section 4 shows how the cumulative distribution of SNRs with diameter may be
interpreted within our framework, and how we can explain the puzzling
linear trend of this distribution without requiring, as usually done, that
SNRs expand linearly to large sizes.
Section 5 presents our conclusions.


\section{Basic ideas, assumptions, and formulae}
\label{sect:Basic}

Our analysis is based on the fundamental criterion that the observed
correlation in the \SgD\ parameter plane originates from the combined effect
of evolutionary tracks in very different ambient conditions.
In this section, we implement this idea by adding some derived /
secondary assumptions that will allow us to develop a more general scenario,
on which our subsequent statistical analyses will be based.


\subsection{When radio SNRs are preferentially seen}
\label{sect:RadioSeen}

A preliminary consideration is that the conditions
in which a given object is most likely observed, during its evolution,
are those in which it spends most of its time.
Since SNR expansion decelerates during most of their lifetime, it is
statistically more likely to find them when their size is close to its
final value.
We are interested in finding SNRs that are visible in radio.
Therefore, it is more important, in this case, to determine the evolutionary
stage at which
the processes responsible for enhancing magnetic fields and/or for producing
high energy electrons are no longer efficient.
In the following, we refer to this phase as the ``final stage'' of a
radio SNR, but it should be clear that it is not the maximum size that a
SNR can reach dynamically, before merging into the ambient medium.

We parametrize the SNR expansion by a power law ($D\propto t^{1/a}$;
$a>1$) up to a maximum size ($D_2$) beyond which the SNR is no longer
detectable in radio.
During its evolution of a given SNR as a radio source,
the probability of
being observed with a given size $D$ is proportional to $dt/dD$, namely
\begin{equation}
\label{eq:probD}
  \prob(D)=aD^{a-1}/D_2^a,\quad\hbox{\rm where}\quad D<D_2,
\end{equation}
(the initial diameter of this evolutionary phase, $D_1$, not being relevant
provided that $(D_1/D_2)^{a-1}\ll1$), so that the average value and standard
deviation of the (decimal) logarithm of $D$ are

\begin{eqnarray}
\label{eq:averlgD}
  \avg{\lg D}&=&\lg D_2-\frac{1}{a\ln 10},                                   \\
\label{eq:sigmalgD}
  \sg_{\lg D}&=&\frac{1}{a\ln 10}.
\end{eqnarray}
For instance, during the adiabatic (Sedov) phase $a=5/2$, so that $\avg{\lg
D}=\lg D_2-0.17$ and $\sg_{\lg D}=0.17$, while, in the later radiative
(pressure-driven snowplow) phase, $a=7/2$, so that $\avg{\lg D}=\lg D_2-0.12$
and $\sg_{\lg D}=0.12$.
This means that, on average, SNR diameters should be rather close to
$D_2$, and that their dispersion should be rather small,
i.e., an individual SNR, during its evolution, is seen to migrate only
slightly in the \SgD\ parameter plane.
For this reason, we propose that selection effects, while important to
determining
the overall distribution of points across the \SgD\ plane, should only have
a marginal effect on the observed probability $\prob(D)$.


\subsection{The end of the radio phase}
\label{sect:EndRadio}

We consider, in particular, the end of the Sedov phase.
According to Truelove \& McKee (\cite{tm99}), it should correspond to a size
\begin{equation}
\label{eq:DB}
  D_B\sim28\,\left(\ESN/10^{51}\U{erg}\right)^{2/7}\nz^{-3/7}\U{pc},
\end{equation}
(where $\ESN$ is the energy of the supernova explosion), while the dynamical
end of the SNR, namely where it merges with the ambient medium, can be placed
at much larger sizes.
The above formula has been obtained by approximating the plasma cooling
function with a power law $\Lmb\propto T^{-1/2}$, where $T$ is the gas
temperature (while other papers use different power-law approximations;
for instance, Blondin et al.\ \cite{bea98} use $\Lmb\propto T^{-1}$).
For a generic $\Lmb\propto T^{-\al}$ relation one may find that
\begin{equation}
\label{eq:DBgeneric}
  D_B\propto\ESN^{(3+2\al)/(11+6\al)}\nz^{-(5+2\al)/(11+6\al)}.
\end{equation}
For $\al$ changing from $1/2$ to $1$, the exponent of $\nz$ in the above
formula changes from 0.429 to 0.412: namely, the numerical value of that
exponent is very weakly dependent on the power-law approximation used.
In the following, we shall then use, without loss of generality, the formula
(here Eq.~\ref{eq:DB}) of Truelove \& McKee (\cite{tm99}).

The correlation found by Berkhuijsen between $D$ and $\nz$ (Eq.\
\ref{eq:DBerk}) is consistent with
\begin{equation}
\label{eq:DovDB}
  D\simeq0.54\left(\ESN/10^{51}\U{erg}\right)^{-2/7}D_B,
\end{equation}
namely with a $D/D_B$ ratio that is independent of $\nz$.

\begin{figure}
\resizebox{\hsize}{!}{\includegraphics{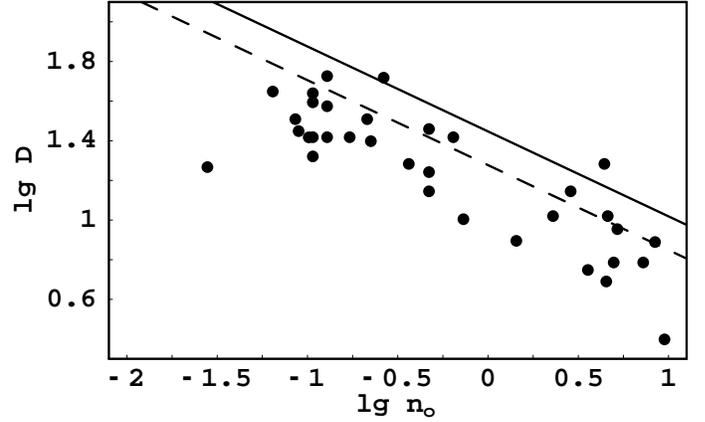}}
\caption{Distribution of SNRs (from Berkhuijsen (\cite{b86}) sample) in the
$\lg\nz$--$\lg D$ parameter plane. For comparison, the theoretical line
corresponding to the end of the Sedov phase (as from Truelove \& McKee
\cite{tm99}) is shown (solid line), as well as that of $\avg{\lg D}$, as
from Eq.~\ref{eq:averlgD} (dashed line).}
\label{Fig2}
\end{figure}

This can be also seen from Fig.~\ref{Fig2}, where the data points from
Berkhuijsen (\cite{b86}) are displayed together with $D_B$, as evaluated
for $\ESN=10^{51}\U{erg}$ (solid line): we note that the line is not a fit,
namely there are no free parameters to tune.
This indicates that most of the known radio SNRs are observed close to the end
of the
Sedov phase, and that in general SNRs must extinguish their radio emission
somewhere close to the end of their Sedov phase.
It is then reasonable to use a Sedov law ($a=5/2$) to approximate the expansion
law during the final phases of radio SNRs: therefore, in the following,
whenever a numerical value for $a$ is required, we shall use the Sedov value.
For a Sedov expansion, the dashed line in Fig.~\ref{Fig2} applies to the
value of $\avg{\lg D}$, given by Eq.~\ref{eq:averlgD}.
The best-fit level (Eq.~\ref{eq:DovDB}) is only 20\% lower, providing a good
argument for SNRs being (statistically) mostly visible around
the end of their adiabatic stage.

It remains to be understood for which physical reason the end of the Sedov
phase should roughly correspond to the switching off of the radio emission.
There is also evidence that the extinguishing transient evolution
must be rather rapid.
Otherwise, we should also see SNRs with $\Sigma$ values considerably lower
than that derived from the \SgD\ correlation for the same size; namely,
in the \SgD\ parameter plane, we should have points spread over
the half-plane below the main correlation (of course, limited to the region
of the \SgD\ plane for which one expects SNRs to be detectable).
This latter piece of observational evidence is also not easy to explain.
The underlying problem is that the physical processes behind the injection
of electrons are poorly understood in the general case, and are probably even
harder to model for conditions near to marginal efficiency.


\subsection{The ``final-stage'' approximation}
\label{sect:FinalStage}

We introduce in a parametric form a basic set of
equations to describe the observed correlations and distributions,
and eventually provide some constraints on future physical models of
the injection of electrons in SNRs.
That the \SgD\ empirical relation is a power law suggests (even though
it does not strictly imply) that all formulae of interest can be approximated
by power laws, thus simplifying considerably the treatment.

As a starting point, we consider the extreme approximation that each of
them is observed very close to its final stage as a radio SNR, namely that
each individual evolutionary track in the \SgD\ plane can be assimilated to
just one point, corresponding to its final position ($D_2$, $\Sg_2$).
We also assume that the dependence of both quantities on the
ambient density ($\nz$) can be approximated by the following power laws:
\begin{eqnarray}
\label{eq:D2parfcn}
  D_2(\nz)&=&K_1\nz^m,\\
\label{eq:Sg2parfcn}
  \Sg_2(\nz)&=&K_2\nz^n.
\end{eqnarray}
In this limiting case, the slope of the \SgD\ relation
\begin{equation}
\label{eq:SgDfcn}
  \Sg_2(D_2)=K_3D_2^\xi,
\end{equation}
would simply be $\xi=n/m$.
In the above formulae, the functional dependence on other physical parameters
is not given explicitely; however, other quantities may be involved.
For instance, if SNRs really are efficient radio emitters only until the end
of the Sedov phase (as suggested by Fig.~2), $D_2$ should
also depend on the SN energy (see Eq.~\ref{eq:DBgeneric}).

To describe the distribution of points along the correlation,
one must also introduce the function $\tP(\nz)$, giving the probability of
finding a SNR in a region of a given density: this probability combines
the density distribution of the interstellar medium, the dependence of the
SN rate on the local density, and how the lifetime of a radio SNR depends
on the ambient conditions.
For the sake of simplicity, and in the absence of any observational evidence
against it, we also approximate this function by a power law, namely
\begin{equation}
\label{eq:tpparfcn}
  \tP(\nz)=K_4\nz^w.
\end{equation}
This distribution is used in  Sect.~\ref{sect:Paradox}.


\subsection{Introducing the SNR evolution}
\label{sect:SNevol}

From this point on, we shall remove the ``final-stage'' approximation,
introduced in the previous section.
Nonetheless the evolution of individual SNRs will still be treated in a
very simplified way, by assuming that SNR evolution in different ambient
conditions differ only by a scaling law and, in practice, by only adopting
power-law behaviors.

We introduce the SNR expansion in the following parametric form:
\begin{equation}
\label{eq:expparfcn}
  t(D,\nz)=K_5 D^a\nz^b
\end{equation}
(for instance, in the Sedov case $a=5/2$, $b=1/2$, and $K_5\propto\ESN^{-1/2}$).
As for the evolution in surface brightness, by assuming that a scaling law
holds, the surface brightness could be expressed in a rather general form as
\begin{equation}
\label{eq:sgparfcnb}
  \Sg(D,\nz)=f(D/D_2(\nz))\Sg_2(\nz),
\end{equation}
where $f(x)$ must vanish at $x>1$.
In this paper, we use a power-law approximation
\begin{equation}
\label{eq:sgparfcn}
  \Sg(D,\nz)=\left(\frac{D}{D_2(\nz)}\right)^p\Sg_2(\nz)
    =K_6D^p\nz^q\quad\hbox{\rm for}\quad D<D_2,
\end{equation}
where $K_6=K_2/K_1^p$ and $q=n-mp$.
Parameters $p$ and $q$ in Eq.~\ref{eq:sgparfcn} can be derived independently.
If the data sample is not heavily affected by selection effects, these
parameters can
be evaluated by simply applying a bilinear regression.

Determining the functional dependence of $\Sg(D,\nz)$ would also allow one
to derive the trajectory of individual SNRs in the parameter plane.
They are simply given by the function $\Sg(D)$ for a constant value of $\nz$,
which in principle differs from the $\Sg(D)$ relation, as traditionally
obtained, because the latter relation is obtained by combining cases with
different $\nz$ values.
In the power-law case, the slope of the evolutionary track of an individual
SNR is then given by the exponent $p$.
In the following, we shall provide some evidence that the value of $p$ is
different from that of $\xi$: this means that evolutionary tracks in the \SgD\
parameter plane have a different slope from that of the overall \SgD\ relation.

Also $m$ can be derived, by fitting Eq.~\ref{eq:D2parfcn} to the $D$--$\nz$
data (provided that $\avg{D}$ is a constant fraction of $D_2$, as from
Eq.~\ref{eq:averlgD}).
On the other hand, there is no way of deriving $a$ and $b$ (defined in
Eq.~\ref{eq:expparfcn}) directly from the correlations between $\Sg$,
$D$, and $\nz$.
Coefficient $a$ could be inferred, in principle, only by studying the
distribution of points about the main correlation (a point that will be treated
in a forthcoming paper), while there is no way of estimatig the exponent
$b$, because it would only affect the distribution of points with $\nz$
along the correlations, a piece of information that is already included in the
definition of the distribution $\tP$ (Eq.~\ref{eq:tpparfcn}).

In principle, a distribution of SN energies may also contribute; but, for
the present analysis, we assume that SN energies are not correlated with any
other quantity and therefore that a distribution of energies would just
produce an additional dispersion across the correlations, without affecting
any of the above slopes.
Some correlation could be possible, in principle, if different population
stars have different distributions of their SN energies.
However, to our knowledge no evidence in favour of this has been presented
so far.


\section{Statistical analysis of the data}
\label{sect:StatAnalysis}


\subsection{Data sample and best-fit parameters}
\label{sect:SampleFit}

We now apply the analysis outlined above to the data
published by Berkhuijsen (\cite{b86}).
That paper
presents a fundamental work on the subject and, although since then a great
number of surveys of higher accuracy
have been performed, it still contains the most extended
data sample of SNRs in which, in addition to the SNR radio surface brightness
and size, quantities derived by other spectral bands are also tabulated.
We extracted from this data sample all SNRs with available data on $\Sg$,
$D$ (radio), and $\nz$ as well.
We excluded SN~1006, because it is now known that the bulk of its
X-ray emission is non-thermal.
The total number of selected objects is 34: the original data sample is given
in Table~1 (for two objects, Cas~A and Tycho, in which two different values
of $\nz$ are given, we took their geometrical mean).
Since distance estimates for Galactic SNRs have changed with time, in the
last column of Table~1 we list the SNR sizes obtained from the most recent
version of the Galactic SNR catalog, by Green (\cite{g09}): they differ
substantially from Berkhuijsen's values only for Kepler and Vela.
The estimated distances of the Large Magellanic Cloud (LMC)
and the Small Magellanic Cloud (SMC) have also slightly changed, from 55~kpc
and 63~kpc (as in Berkhuijsen \cite{b86}), to 48~kpc and 61~kpc (Macri et al.\
\cite{mea06}, Hilditch et al.\  \cite{hea05}), respectively.
In our calculations, we used all of these new distances, and we revised
accordingly the SNR linear size and density estimates (being $\nz\propto
d^{-1/2}$).
The last two columns of Table~1 provide the $D$ and $\nz$ values that we used.

\begin{table}
\caption{The data sample (extracted from Berkhuijsen 1986)}
\label{table:1}
\centering
\begin{tabular}{r r c c c c c}
\hline\hline
& Object & $-\lg\Sigma$ & $D$ & $\nz$ & $D$ & $\nz$ \\
& &
  & \multicolumn{2}{c}{(as from}
  & \multicolumn{2}{c}{(after distance} \\
& & $\U{(W\,m^{-2}}$
  & \multicolumn{2}{c}{Berkhuijsen)}
  & \multicolumn{2}{c}{revision)} \\
& & $\U{Hz^{-1}sr^{-1})}$ & $\Ub{pc}$ & $\Ub{cm^{-3}}$
                          & $\Ub{pc}$ & $\Ub{cm^{-3}}$ \\
\hline
GAL & W44       & 19.409 & 26.0	& 0.22 & 25.0 & 0.22 \\
GAL & Cas A     & 16.721 &  4.1 & 4.94 &  4.9 & 4.52 \\
GAL & Tycho     & 18.921 &  5.4 & 3.64 &  5.6 & 3.57 \\
GAL & RCW103    & 19.377 &  9.6 & 5.04 &  9.0 & 5.21 \\
GAL & Kepler    & 18.481 &  3.8 & 7.70 &  2.5 & 9.49 \\
GAL & W49B      & 18.432 & 11.0 & 0.70 & 10.1 & 0.73 \\
GAL & VelaXYZ   & 20.367 & 36.0 & 0.02 & 18.5 & 0.03 \\
GAL & RCW86     & 20.276 & 35.0 & 0.08 & 28.1 & 0.09 \\
LMC & 0453--685 & 19.730 & 20.0 & 0.44 & 17.5 & 0.47 \\
LMC & 0454--665 & 19.629 & 16.0 & 0.44 & 14.0 & 0.47 \\
LMC & 0455--687 & 20.299 & 51.0 & 0.06 & 44.5 & 0.06 \\
LMC & 0500--702 & 20.590 & 30.0 & 0.12 & 26.2 & 0.13 \\
LMC & 0505--679 & 20.374 &  9.0 & 1.34 &  7.9 & 1.43 \\
LMC & 0506--680 & 18.976 & 12.0 & 2.14 & 10.5 & 2.29 \\
LMC & 0509--675 & 19.231 &  7.0 & 4.64 &  6.1 & 4.97 \\
LMC & 0519--697 & 19.524 & 24.0 & 0.10 & 20.9 & 0.11 \\
LMC & 0519--690 & 18.984 &  7.0 & 6.76 &  6.1 & 7.24 \\
LMC & 0520--694 & 20.435 & 37.0 & 0.20 & 32.3 & 0.21 \\
LMC & 0525--660 & 19.480 & 30.0 & 0.60 & 26.2 & 0.64 \\
LMC & 0525--696 & 18.293 & 22.0 & 4.12 & 19.2 & 4.41 \\
LMC & 0525--661 & 18.544 & 16.0 & 2.68 & 14.0 & 2.87 \\
LMC & 0527--658 & 20.807 & 61.0 & 0.12 & 53.2 & 0.13 \\
LMC & 0528--692 & 20.168 & 30.0 & 0.16 & 26.2 & 0.17 \\
LMC & 0532--710 & 19.947 & 43.0 & 0.12 & 37.5 & 0.13 \\
LMC & 0534--699 & 20.356 & 33.0 & 0.44 & 28.8 & 0.47 \\
LMC & 0534--705 & 20.481 & 37.0 & 0.08 & 32.3 & 0.09 \\
LMC & 0535--660 & 18.287 & 12.0 & 4.30 & 10.5 & 4.60 \\
LMC & 0536--706 & 20.481 & 30.0 & 0.10 & 26.2 & 0.11 \\
LMC & 0543--689 & 20.758 & 45.0 & 0.10 & 39.3 & 0.11 \\
LMC & 0547--697 & 19.736 & 50.0 & 0.10 & 43.6 & 0.11 \\
LMC & 0548--704 & 20.151 & 22.0 & 0.34 & 19.2 & 0.36 \\
SMC & 0045--734 & 19.802 & 27.0 & 0.10 & 26.1 & 0.10 \\
SMC & 0102--722 & 18.570 &  8.0 & 8.28 &  7.7 & 8.41 \\
SMC & 0103--726 & 20.807 & 54.0 & 0.26 & 52.3 & 0.26 \\
\hline
\end{tabular}
\end{table}

The average ambient density is estimated by Berkhuijsen (\cite{b86}) in a
simple way, using the following relation (derived from Long \cite{l83}):
\begin{equation}
\label{eq:nzformula}
  \nz=(6/\pi)^{1/2}\eps^{-1/2}f^{1/2}\LX^{1/2}D^{-3/2},
\end{equation}
where $\eps$ (taken to be $3\E{-23}\U{erg\,cm^{-3}\,s^{-1}}$) is the specific
emissivity, $f$ is the filling factor (taken to be close to unity), and
$\LX$ is the X-ray luminosity.
While the exact values of $\eps$ and $f$ are not important to our statistical
analysis, it is crucial that these quantities remain constant, or at least
independent of other parameters (such as size and surface brightness).
In spite of its simplicity, this formula provides reasonably good results.
Upper limits to the uncertainty in this $\nz$ evaluation can be derived from
the dispersion about the $\lg D$--$\lg\nz$ regression.
Based on the assumption that the measured dispersion depends only on the
uncertainties in
$\nz$, one obtains $\sg(\lg\nz)=0.42$; while also taking into account
the dispersion in $D$, as modeled by Eq.~\ref{eq:sigmalgD} for the case of
Sedov expansion, one derives a residual dispersion $\sg(\lg\nz)=0.18$, namely
a typical uncertainty in the density derived by Berkhuijsen of only about 50\%.

\begin{figure}
\resizebox{\hsize}{!}{\includegraphics{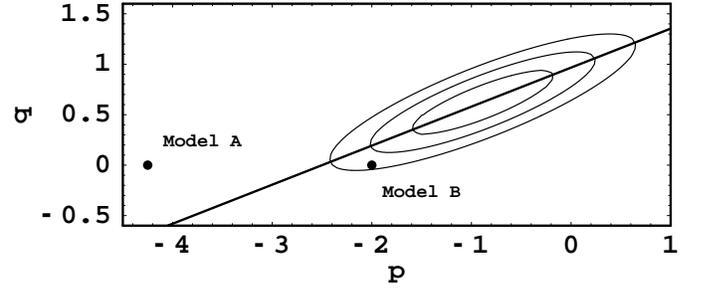}}
\caption{Plot of the confidence levels in the $p$-$q$ parameter plane.
The levels plotted correspond to 1, 2, and 3-$\sg$ confidence levels, while the
line indicates the maximum spread direction (see Eq.~\ref{eq:princcomp}).
For comparison, two theoretical predictions are plotted: ``Model A''
($-17/4,0$) indicates the case with constant efficiency in both particle
acceleration and magnetic field compression plus amplification (Berezhko \&
V\"olk \cite{bv04}); while ``Model B'' ($-2,0$) refers to the case in which
particles are accelerated with constant efficiency but the magnetic field
is constant (see text).}
\label{Fig3}
\end{figure}

By performing linear regressions between the logarithmic quantities (which
is equivalent to assuming constant relative errors in the measurements), we
obtain:
\begin{eqnarray}
\label{eq:bestfitm}
  m&=&-0.37\pm0.04, \\
\label{eq:bestfitxi}
  \xi&=&-2.06\pm0.34, \\
\label{eq:bestfitp}
  p&=&-0.89\pm0.57, \\
\label{eq:bestfitq}
  q&=&+0.62\pm0.25,
\end{eqnarray}
(where 1-$\sg$ uncertainties are indicated).
It is apparent that, while $m$ and $\xi$ are found to have reasonably
small uncertainties, the uncertainties of $p$ and $q$ are larger.
The reason is that, in the data, there is a near-degeneracy between $p$ and
$q$, as is well shown by a plot of the confidence levels (Fig.~\ref{Fig3}).
In this sense, a combined quantity that can be far more reliably determined is:
\begin{equation}
\label{eq:princcomp}
  q-0.39p=0.97\pm0.14.
\end{equation}

A potential problem of this sample, and of SNR samples in general, is the
presence of selection effects.
In the introduction, we mentioned the analyses of Green
(\cite{g05}) and Uro\v sevi\'c et al.\ (\cite{uea05}) on this subject.
The points raised by Green (\cite{g05}) are more appropriate to our sample,
which consists only of SNRs located in our Galaxy and the Magellanic Clouds.

\begin{figure}
\resizebox{\hsize}{!}{\includegraphics{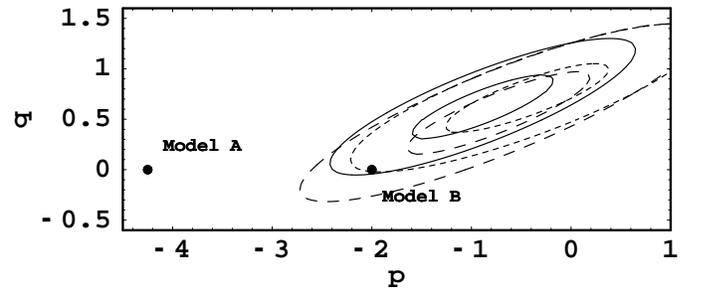}}
\caption{
Plot of the 1-$\sg$ and 3--$\sg$ confidence levels in the $p$-$q$ parameter
plane, for the whole sample (solid lines), as well as for selections of
the 25 and 30 SNRs with the highest surface brightness (short-dashed and
long-dashed lines, respectively).
}
\label{Fig4}
\end{figure}

Even though a detailed treatment of the selection effects is beyond the
scope of this paper (but will be approached in a forthcoming paper), for
the sake of illustration we repeated the computations that led
to Fig.~\ref{Fig3}, but on subsamples containing, respectively, the 25 and 30
SNRs with the highest radio surface brightness with corresponding thresholds
in $\lg\Sg$ of $-20.356$ and $-20.481$ respectively (see Table~1).
Figure~\ref{Fig4} shows a comparison of the confidence levels (only 1-$\sg$
and 3--$\sg$, for figure clarity) for the two subsamples, superimposed on those
for the entire sample (as in Fig.~\ref{Fig3}).
It is apparent that any additional selection in surface brightness does not
generate any substantial difference in the results, apart from a slight
terms of
broadening compatible with the lower sample size being analyzed by statistics
and a slight
shift toward lower $p$ and $q$ values (the latter point will be discussed
in the next section).

\subsection{Testing the ``constant efficiencies'' model}
\label{sect:ConstEff}

Among the various theoretical attempts to model the radio emission from SNRs,
one of the most recent and popular is that by Berezhko \& V\"olk (\cite{bv04}).
This paper assumes that the kinetic energy density entering the shock
($m\nz\Vsh^2$, where $m$ is the mean atomic mass) is converted with constant
efficiencies into the energy densities of magnetic field and accelerated
electrons ($\epsB$ and $\epsCR$, respectively): this means, for instance,
that the effective magnetic field in the synchrotron emitting region decreases
with time, simply because the SNR shock is slowing down.
For synchrotron emission with a power index $-0.5$ (namely the average index
for radio SNRs), the surface brightness should scale as
\begin{equation}
\label{eq:sigmasynch}
  \Sg\propto KB^{3/2}D\propto\left(\epsCR\epsB^{3/4}\right)
    D\propto\left(\nz\Vsh^2\right)^{7/4}D
\end{equation}
A further assumption of this model is Sedov expansion, which implies that
$m\nz\Vsh^2\sim\ESN/D^3$, so that one finally obtains
\begin{equation}
\label{eq:sigmaBV}
  \Sg\propto\ESN^{7/4}D^{-17/4},
\end{equation}
i.e., with $p=-4.25$, and $q=0$ (labeled as ``Model A'' in Fig.~\ref{Fig3}).
Namely, according to this model, individual SNR tracks in the \SgD\ plane
must be rather steep and ``independent of the ambient density''.
Since Berezhko \& V\"olk (\cite{bv04}) state that the slope of the \SgD\
relation should represent the slope of individual evolutionary tracks,
they predict that $\xi=-4.25$ should be the slope of the \SgD\ relation.
However, neither their predicted value for $\xi$ matches the data, nor
does (and at an even higher significance level) their predicted $(p,q)$ pair
(see Fig.~\ref{Fig3}).
It is unlikely that this mismatch is a mere consequence of a sample
incompleteness in surface brightness.
Figure~\ref{Fig4} shows that for subsamples in which a further
selection in surface brightness has been applied the barycenter of the
confidence levels moves only mildly.

A more appropriate model (in the sense that it is ``only'' about 2-$\sg$ away
from the best-fit model values) would be one in which electrons are accelerated
with constant efficiency ($\epsCR\propto\nz\Vsh^2$) but the magnetic field
is taken to be constant, not only during the evolution of an individual SNR
but also among different SNRs.
This happens, for instance, if the post-shock field has been compressed only
by the shock, i.e., is proportional to the ambient field, which in turn is
roughly constant (see e.g., Crutcher et al.\ \cite{cht03}), in near
equipartition with the interstellar thermal pressure of the diffuse
interstellar medium.
This case does not exclude the presence of an extra field amplification,
provided that it yields a constant factor.
As in the previous case, but assuming $B$ to be constant, one can now write
\begin{equation}
\label{eq:sigmasynchB}
  \Sg\propto KD\propto\epsCR D\propto\left(\nz\Vsh^2\right)D\propto
  \ESN D^{-2},
\end{equation}
namely $p=-2$ and $q=0$ (which is labeled ``Model B'' in Fig.~\ref{Fig3}).


\subsection{The results with a more ``physical'' flavour}
\label{sect:PhysFlavour}

\begin{figure}
\resizebox{\hsize}{!}{\includegraphics{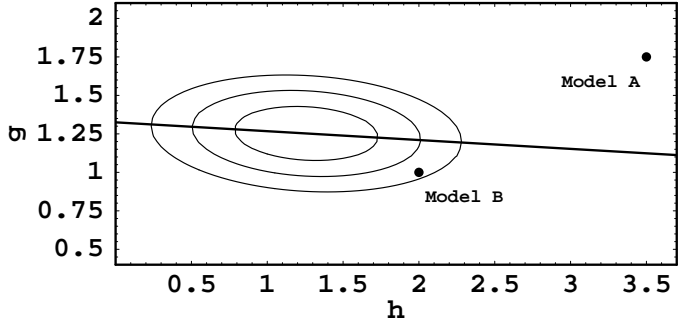}}
\caption{Same as Fig.~\ref{Fig3}, in the $g$-$h$ parameter plane.}
\label{Fig5}
\end{figure}

We can approach the problem from the opposite direction, by trying to translate
the information derived in terms of $(p,q)$ into constraints on the physics
that controls the magnetic and cosmic-ray efficiencies.
We assume that
\begin{equation}
\label{eq:physicpar}
  KB^{3/2}\propto\nz^g\Vsh^h,
\end{equation}
where $g$ and $h$ are free parameters.
This is not the most general case, but it simply relates the efficiencies
to primary local quantities encountered by the shock.
Based on this assumption, and an expansion law $t\propto D^a$ passing through
the endpoint of the Sedov phase $(D_B,t_B)$, the equation for the surface
brightness becomes
\begin{equation}
\label{eq:sigmapar}
  \Sg\propto\ESN^{(4a-3)/14}D^{1-(a-1)h}\nz^{g-(3a-4)h/7}.
\end{equation}
In the case of Sedov expansion, this equation simplifies into
\begin{equation}
\label{eq:sigmaparsed}
  \Sg\propto\ESN^{1/2}D^{1-3h/2}\nz^{g-h/2}.
\end{equation}
Figure~\ref{Fig5} shows the confidence levels in this new pair of parameters.
An advantage is that the direction of maximum dispersion is almost parallel
to the $h$ axis, which means that at least the best-fit value for $g$ is
well determined.
We have
\begin{eqnarray}
\label{eq:bestfitg}
  g&=&1.25\pm0.14, \\
\label{eq:bestfith}
  h&=&1.26\pm0.38.
\end{eqnarray}
For comparison, the ``constant efficiencies'' model prescribes that $g=1.75$
and $h=3.5$, as can easily be derived from Eqs.~\ref{eq:sigmaBV} and
\ref{eq:sigmaparsed}.
What we have found here is not different from the previous section, but is
simply
displayed in a more physical way.

To summarize, in most SNRs the constant efficiency assumption (namely for
both field amplification and particle acceleration) does not hold.
This may not be surprising, in the view that the statistically most common
cases are those of SNRs close to their radio endpoint, namely when particle
acceleration is close to being halted.
Using observations to test these critical cases may indeed be important
to obtaining a clearer
understanding of the physical processes responsible for magnetic
amplification and particle acceleration.


\section{Results from an independent sample: M~33}
\label{sect:M33}

As already mentioned in the introduction, SNR samples studied in other
galaxies provide promising input to this kind of analysis.
On the other hand, the presence of selection effects could affect the results
(Uro\v sevi\'c et al.\ \cite{uea05}), and a statistical analysis of these
samples should include a careful study of these.
Extragalactic SNR samples will be considered in more detail in a forthcoming
paper. For mere comparison with what has already been obtained using the
sample of Berkhuijsen, we present here the results for a M~33 SNR data
sample, which is a sample completely independent from that used so far.

The sample was obtained by selecting SNRs for which radio fluxes are given by
Gordon et al.\ (\cite{gea99}), X-ray fluxes by Pietsch et al.\ (\cite{pea04}),
and (optical) diameters by Gordon et al.\ (\cite{gea98}).
The X-ray survey of Plucinsky et al.\ (\cite{pea08}), based on Chandra data,
was also used to solve some cases of uncertain identification.
In this way, we selected 22 SNRs: the data sample is shown in
Table~\ref{table:2}.
The first 3 columns report the SNR identification numbers in the various
catalogs, respectively, Gordon et al.\ (\cite{gea98}; labeled by ``opt'' in the
Table), Gordon et al.\
(\cite{gea99}; labeled by ``rad''), and Pietsch et al.\ (\cite{pea04}; labeled
by ``xray''), while the
next 3 columns show, respectively, the published linear (optical) diameters,
20~cm fluxes, and measured (i.e., absorbed) 0.2--4.5~keV fluxes.

\begin{table}
\caption{M~33 data sample}
\label{table:2}
\centering
\begin{tabular}{r r r  c c c}
\hline\hline
ID & ID & ID & D & S(20 cm) & Flux(0.2--4.5) \\
opt & rad & xray & $\Ub{pc}$ & $\Ub{mJy}$ & $\Ub{\flux}$ \\
\hline
 9 &  11 &  93 & 18 & 0.7 & $4.29\E{-15}$ \\
11 &  13 &  98 & 17 & 0.6 & $3.81\E{-15}$ \\
15 &  20 & 106 & 27 & 0.6 & $2.79\E{-15}$ \\
20 &  25 & 120 & 10 & 0.8 & $1.03\E{-14}$ \\
21 &  29 & 121 & 28 & 0.9 & $1.41\E{-13}$ \\
25 &  42 & 144 & 27 & 1.4 & $3.63\E{-15}$ \\
27 &  47 & 153 & 23 & 1.2 & $2.99\E{-15}$ \\
28 &  50 & 158 & 11 & 0.8 & $2.45\E{-14}$ \\
29 &  52 & 161 & 20 & 0.5 & $1.73\E{-14}$ \\
31 &  57 & 164 & 39 & 1.8 & $3.88\E{-14}$ \\
35 &  64 & 179 & 32 & 3.5 & $1.01\E{-14}$ \\
42 &  75 & 194 & 29 & 0.5 & $1.39\E{-14}$ \\
47 &  90 & 207 & 36 & 0.2 & $7.05\E{-15}$ \\
53 & 110 & 213 & 40 & 0.2 & $1.53\E{-15}$ \\
54 & 111 & 214 & 16 & 1.3 & $2.08\E{-15}$ \\
55 & 112 & 215 & 18 & 4.4 & $2.84\E{-14}$ \\
57 & 114 & 220 & 21 & 0.4 & $1.87\E{-15}$ \\
59 & 121 & 224 & 16 & 0.3 & $3.90\E{-15}$ \\
62 & 125 & 225 & 29 & 0.4 & $4.87\E{-15}$ \\
64 & 130 & 230 & 27 & 0.5 & $3.65\E{-15}$ \\
73 & 148 & 250 & 17 & 0.5 & $1.41\E{-14}$ \\
97 & 181 & 314 & 35 & 0.8 & $4.84\E{-15}$ \\
\hline
\end{tabular}
\end{table}

To derive 1~GHz radio fluxes, we extrapolated the 20~cm fluxes 
tabulated by Gordon et al.\ (\cite{gea99}), by assuming a spectral index of
$-0.5$.
For most SNRs, Gordon et al.\ (\cite{gea99}) also estimate spectral indices,
but the uncertainty in these estimates is rather large, and we therefore
preferred to adopt a ``standard'' value for the spectral index.
We evaluated the unabsorbed X-ray fluxes by taking a column density
$N_H=1.0\E{21}\U{cm^{-2}}$ towards M~33 (Plucinsky et al.\ \cite{pea08}), and
assuming for the average SNR spectrum a Raymond-Smith model with
$kT=0.3\U{keV}$.
Using WebPIMMS
\footnote{{\tt http://heasarc.gsfc.nasa.gov/Tools/w3pimms.html}},
a correction factor of $\sim1.94$ is evaluated: the precise value of this
factor
is not very important for our purposes, provided that the X-ray SNR spectra
are not too different among themselves.
Finally, a M~33 distance of 817~kpc (Freedman et al.\ \cite{fea01}) is assumed
here; since Gordon et al.\ (\cite{gea98}) use a distance of 840~kpc, for
consistency we applied a small correction to their published SNR sizes.

Here are our results, to be compared with those presented above.
The formulae equivalent to Eqs.~\ref{eq:bestfitm}--\ref{eq:bestfitq},
\ref{eq:bestfitg} and \ref{eq:bestfith} are respectively
\begin{eqnarray}
\label{eq:bestfitm33}
  m&=&-0.34\pm0.07, \\
\label{eq:bestfitxi33}
  \xi&=&-2.20\pm0.46, \\
\label{eq:bestfitp33}
  p&=&-1.37\pm0.64, \\
\label{eq:bestfitq33}
  q&=&+0.52\pm0.30, \\
\label{eq:princcomp33}
  q&-&0.37p=1.04\pm0.20 \\
\label{eq:bestfitg33}
  g&=&1.31\pm0.20 \\
\label{eq:bestfith33}
  h&=&1.58\pm0.42;
\end{eqnarray}
while Figs.\ 6, 7, 8, and 9 correspond, for M~33, to Figs.\ 1, 2, 3, and
5, respectively.
It is apparent that all results from this further sample show close
agreement, within the quoted errors, with what we found above using
the data of Berkhuijsen.
A comparison of the regression results for the two samples is given
in Table~3.

\begin{figure}
\resizebox{\hsize}{!}{\includegraphics{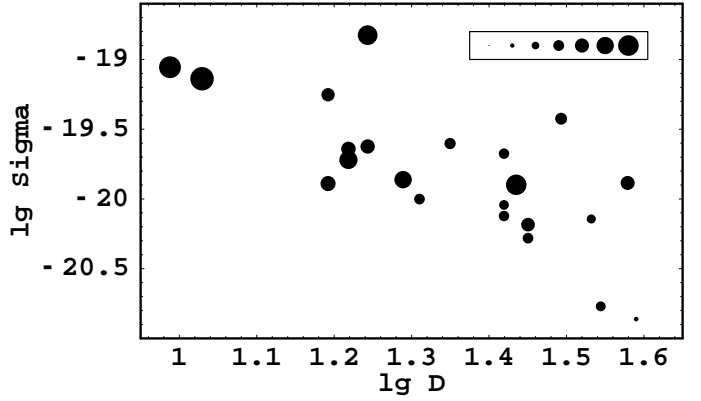}}
\caption{
Distribution of the M~33 SNRs sample in the $\lg D$--$\lg\Sg$
parameter plane (to be compared with Fig.~1).
The dot sizes are proportional to $\lg\nz$ values, the legend showing in
order sizes corresponding to $\lg\nz$ from --1.0 to 0.5, in steps of 0.25.
}
\label{Fig6}
\end{figure}

\begin{figure}
\resizebox{\hsize}{!}{\includegraphics{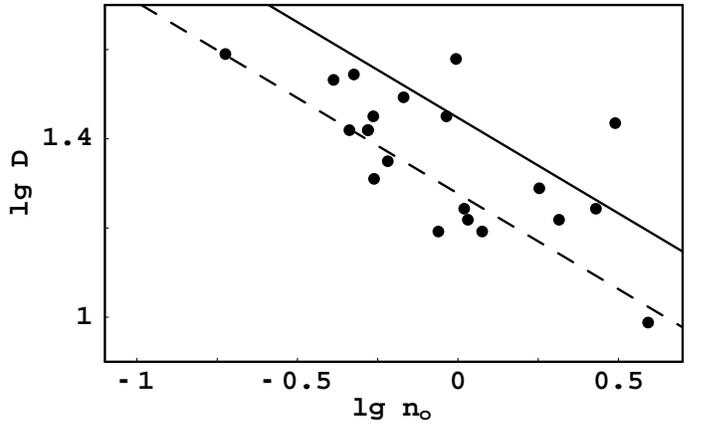}}
\caption{
Distribution of M~33 SNRs in the $\lg\nz$--$\lg D$ parameter plane
(to be compared with Fig.~2).}
\label{Fig7}
\end{figure}

\begin{figure}
\resizebox{\hsize}{!}{\includegraphics{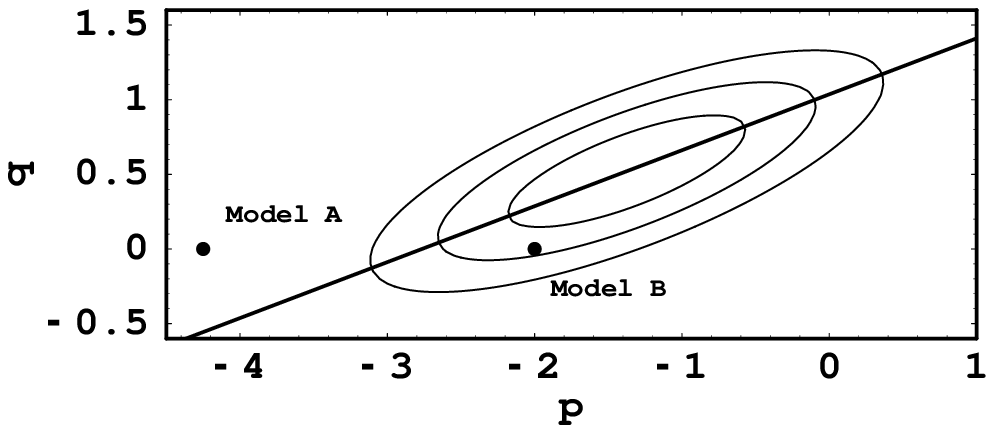}}
\caption{
Plot, for M~33 SNRs, of the confidence levels in the $p$-$q$ parameter plane
(to be compared with Fig.~3).}
\label{Fig8}
\end{figure}

\begin{figure}
\resizebox{\hsize}{!}{\includegraphics{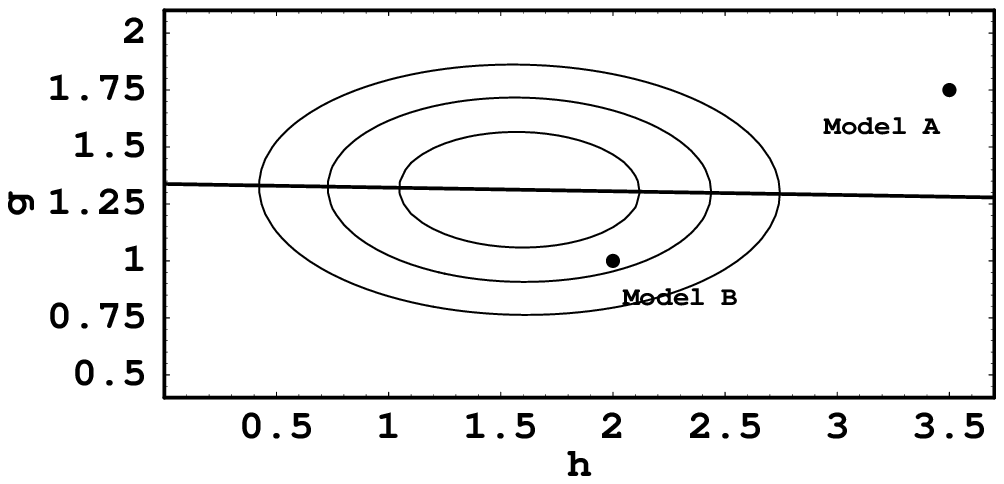}}
\caption{{
Plot, for M~33 SNRs, of the confidence levels in the $g$-$h$ parameter plane (to
be compared with Fig.~5).}}
\label{Fig9}
\end{figure}

\begin{table}
\caption{Synoptic table of the results of the regression analyses}
\label{table:3}
\centering
\begin{tabular}{c c c c c}
\hline\hline
Formula  & &  \multicolumn{2}{c}{Data Samples} \\
                               &        & Berkhuijsen    & M~33 \\
\hline
$D_2(\nz)\propto\nz^m$         & $m=$   & $-0.37\pm0.04$ & $-0.34\pm0.07$ \\
$\Sg_2(D_2)\propto D_2^\xi$    & $\xi=$ & $-2.06\pm0.34$ & $-2.20\pm0.46$ \\
$\Sg(D,\nz)\propto D^p\nz^q$   & $p=$   & $-0.89\pm0.57$ & $-1.37\pm0.64$ \\
                               & $q=$   & $+0.62\pm0.25$ & $+0.52\pm0.30$ \\
$q-\lmb p=\mu$                 & $\lmb=$& $+0.39$        & $+0.37$        \\
                               & $\mu=$ & $+0.97\pm0.14$ & $+1.04\pm0.20$ \\
$KB^{3/2}\propto\nz^g\Vsh^h$   & $g=$   & $+1.25\pm0.14$ & $+1.31\pm0.20$ \\
                               & $h=$   & $+1.26\pm0.38$ & $+1.58\pm0.42$ \\
\hline
\end{tabular}
\end{table}


\section{The SNR cumulative distribution with size}
\label{sect:CumulDistr}


\subsection{The original paradox and how it can be solved}
\label{sect:Paradox}

\begin{figure}
\resizebox{\hsize}{!}{\includegraphics{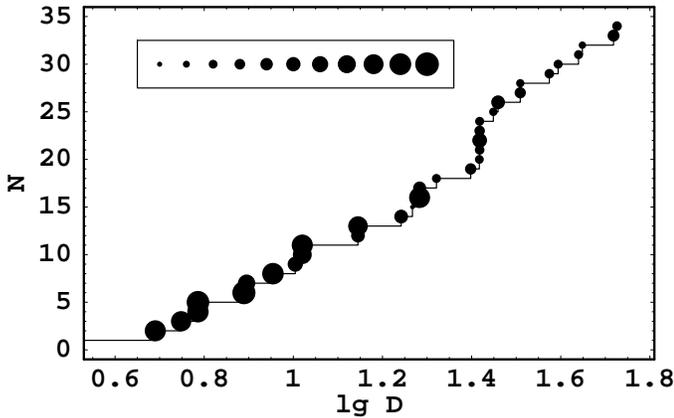}}
\caption{Cumulative distribution, for Berkhuijsen's sample, where the dot
sizes are proportional to the $\lg\nz$ value of the last SNR entering in
the cumulative.
The legenda uses for the dot sizes the same convention as in Fig.~\ref{Fig1}.
}
\label{Fig10}
\end{figure}

The cumulative distribution of the number of SNRs with sizes smaller than a
given diameter ($N$--$D$ relation) is another statistical distribution that
has traditionally been studied.

For the Magellanic Clouds, Mills at al.\ (\cite{mills84}) derived an
almost linear relation ($N\propto D^{1.2}$) up to sizes as large as 40~pc.
They also argued that a similar relation should be present in our Galaxy.
A linear cumulative distribution is usually taken as evidence that these
SNRs are still in free expansion.
However, when the SNR diameter is 40~pc, the swept-up mass is
$\sim(1000/\nz)\U{\Msun}$.
Therefore, except for cases of exceptionally low ambient density,
at those sizes SNRs should already be in the Sedov phase, and therefore
strongly decelerated.
A way to solve this paradox was suggested by Green
(\cite{g84}), which invokes a major effect of sample incompleteness.
Indeed, any sample incompleteness is expected to substantially affect the
cumulative distribution.
Much less pronounced effects on the \SgD\ correlation, as well as on the
correlations discussed in Sect.~\ref{sect:StatAnalysis}, are expected,
since the main effect of the sample incompleteness should be to change the
distribution of points ``along'' the correlations.
The relevance of data sample incompleteness to the correlations will be
discussed in a forthcoming paper.

Within our framework, the cumulative distribution with size is independent
of the expansion law of individual SNRs, but is related instead to the
statistical distribution of the ambient medium density, as defined by
Eq.~\ref{eq:tpparfcn}.
Indeed
\begin{equation}
\label{eq:cumdistr}
  N(D)\propto\tP(\nz)\nz\propto D^{(1+w)/m}.
\end{equation}
In this sense, it is not even necessary to account for
the sample incompleteness,
as done by Green (\cite{g84}).

For the sake of illustration, Fig.~\ref{Fig10} shows the $N(D)$ cumulative
distribution of Berkhuijsen's sample.
Here the close to linearity of the distribution is coincidental, since
there is no reason to expect the sample of Berkhuijsen to be complete.
Instead, the trend of dot sizes with SNR diameter (which is to some extent
a different way of displaying the information contained in Fig.~\ref{Fig2})
clearly shows how, for increasing size, the cumulative is more and more
populated by SNRs located in more tenuous ambient media.


\subsection{The case of M~82}
\label{sect:M82}

Samples of SNRs in nearby galaxies have become increasingly
available, and the close to linearity of the $N$--$D$ relation is a rather
standard property of these samples.

A particularly interesting case is that of M~82.
In this nearby starburst galaxy, a number of radio sources have been detected
(Kronberg et al.\ \cite{kbs85}), which may be SNRs that are much brighter and
much smaller in size (a few parsec at most) than usual.
Their positions on the \SgD\ parameter plane are in all cases
consistent with the
extrapolation to lower sizes of the best-fit \SgD\ relation for other galaxies.
Chevalier \& Fransson (\cite{cf01}) proposed that these sources are SNRs
expanding in
a high-density ambient medium (with densities of order of $10^3\U{cm^{-3}}$).

Kronberg et al.\ (\cite{kea00}) also placed upper limits on the
flux density variations in most of the radio sources: approximately 75\%
of these objects are very stable, with a lower limit of $\sim10^3\U{yr}$
to their characteristic radio-emitting lifetimes.
Based on this upper limit, Seaquist \& Stankovic (\cite{ss07})
suggested that they may not be SNRs, but rather cluster wind-driven bubbles.
Their main argument is that the lack of observed time variability is
inconsistent with the estimated ages of these objects.
That is, if they are SNRs in free expansion (with typical velocities of
$\sim10,000\U{km\,s^{-1}}$), their ages should be a few hundred years at
most, while to account for the lack of variability, velocities no greater than
$\sim500\U{km\,s^{-1}}$ are required (Chevalier \& Fransson \cite{cf01}).

A crucial point in this reasoning is the expansion regime of these objects,
if they are indeed SNRs.
A linear expansion is argued by both Muxlow et al.\ (\cite{mea94}) and Fenech
et al.\ (\cite{fea08}) based on the cumulative distribution with
size being almost linear.
However, as we have explained above, a linear cumulative distribution does not
imply a linear expansion, if it is caused by the combination of SNRs expanding
in different ambient densities.
Indeed, the data for SNRs in M~82 agree well with the extrapolation of
the \SgD\ relation derived for SNRs in other galaxies to smaller sizes.
Thus, the arguments we have exposed in this paper should also be applied to
SNRs in M~82.

Even though they have small sizes in the parsec range (Muxlow et al.\
\cite{mea94}), they could be close to the end of their Sedov phase provided
that the ambient densities are $\sim10^3\U{cm^{-3}}$, with corresponding
shock velocities of $\sim10^3\U{km\,s^{-1}}$ compared to
$\sim10^4\U{km\,s^{-1}}$
as in the case of undecelerated expansion.
These lower shock velocities infer characteristic times of $\sim10^3\U{yr}$,
compatible with the average radio-emitting lifetime found by
Kronberg et al.\ (\cite{kea00}).
The only exceptions are a few fast-evolving radio sources that
are probably radio SNe, namely young objects still evolving in their
circumstellar medium.
Although the detailed physical conditions and processes at such high
densities may differ from those at densities typical of other galaxies,
we are confident that the simple estimate presented above is adequate to
justify the large measured characteristic times within our framework.
Finally, Fenech et al.\ (\cite{fea08}) claimed to have measured very high
expansion velocities, up to $10,000\U{km\,s^{-1}}$: these measurements seem
to conflict not only with our model but also with the above-mentioned
photometric measurements (Kronberg et al.\ \cite{kea00}).
Therefore, it is probably too early to discuss these latest measurements.


\section{Conclusions}
\label{sect:Conclus}

We have shown that studies of the statistical properties of
SNR samples may provide insight into the physics of electron
acceleration and the time evolution of SNRs.
We have proposed a new scenario, along the lines of previous work by
Berkhuijsen (\cite{b86}), which interprets in a natural way
the observed correlations between radio surface brightness $\Sg$, size $D$,
and ambient density $\nz$ in a sample of SNRs.
The main parameter of SNR evolution that enters into these correlations
is the time at which a SNR ceases to behave as a radio source, and we find
that this endpoint is located close to the end of the Sedov phase.
We otherwise find that the observed correlations mostly reflect that the
sample consists of SNRs located in very different ambient conditions;
while the evolution of individual SNRs plays a secondary role, and cannot
be extracted by simply studying correlations between pair of quantities.

Within this framework, we present a new approach to analyzing the statistical
data, based on a 2-dimensional fit to $\Sg$ as a function of $D$ and $\nz$.
We show that the slope of $\Sg(D)$ at constant $\nz$ should represent more
closely
the true evolutionary track of an individual SNR than the well known ``\SgD\
relation'', which is obtained without including information about $\nz$.

In this work, we have used data published by Berkhuijsen (\cite{b86}).
Although this data sample is rather limited, our method of analysis applied
to these data is already capable of discriminating
to some level between different
theoretical models.
For instance, models prescribing constant efficiencies for both magnetic
field (turbulent) amplification and electron acceleration (e.g., Berezhko \&
V\"olk \cite{bv04}) are well outside the parameters region allowed by the data.
On the other hand, models assuming a constant acceleration efficiency but a
constant post-shock magnetic field are marginally (about 2-$\sg$) consistent
with the data.
For the sake of comparison, we have applied the same technique to a sample
of SNRs in M~33.
Although this sample could be affected by selection effects, it is completely
independent from the sample of Berkhuijsen, and the parameters that we derive
from the two samples are in close agreement, within the statistical errors.
This may indicate again that our technique is robust andthe assumptions at the
basis of
our statistical analysis are correct.
With the use of larger and more accurate data samples the confidence interval
will become narrower, and the statistical results will be capable of
constraining models more tightly.

In nearby galaxies, deep and detailed surveys of SNRs have
become available in both radio and X-rays.
The number of SNRs in other galaxies has increased considerably, with
reasonably good radio and X-ray flux measurements and in various cases also
information about their angular size.
Although SNRs in external galaxies are more difficult objects to observe
than Galactic SNRs, their distances are well known (since they correspond 
to the distance of the parent galaxy); in addition, selection
effects can be modeled in a more accurate way for a sample of SNRs that
are all at the same distance.
Therefore, a natural extension of the present work is to elaborate our
diagnostic
tools along the guidelines shown in this paper, and to use them to investigate
SNRs samples in various nearby galaxies.
These results will be presented in a forthcoming paper.


\begin{acknowledgements}
O.P. acknowledges the hospitality of INAF - Osservatorio Astrofisico di
Arcetri, where most of this work was carried out.
R.B. acknowledges the hospitality of KITP, Santa Barbara, during the last
phase of revision of the manuscript.
This work was partially supported by the PRIN-MIUR 2006, by the ASI under Grant
ASI-INAF No.\ I/023/05/0, and by the NSF under Grant No.\ PHY05--51164.
\end{acknowledgements}


\end{document}